\documentclass[10pt,twocolumn]{article}
\usepackage[T1]{fontenc}
\usepackage[utf8]{inputenc}
\usepackage{authblk}
\usepackage{amsmath}
\usepackage{booktabs} 
\usepackage{dcolumn}
\usepackage{subfigure}
\usepackage{numprint}
\usepackage{mathtools}
\usepackage{amssymb}
\usepackage{color}
\usepackage{geometry}
\usepackage{chicago}

\title{We are not alone ! (at least, most of us). \\ Homonymy in large scale social groups}
\author{Arthur Charpentier\footnote{Universit\'e de Rennes 1} \& Baptiste Coulmont\footnote{Universit\'e Paris 8 and INED}}
\begin{document}

\maketitle

\section{First and Last Names Ho\-mo\-nyms}

%{\color{red}EST-CE LA BONNE "CLASSIFICATION"?} 
The Western system of identification is based on a first and a last name : the first name is a personal name, the last name is a transmitted family name, often from the father to his children. According to \citeNP{Scott02} this system is first of all a government device, to monitor individuals and to ensure the rights and duties of citizen : it surfaced with the emergence of state governments. Nowadays, the more stable the state, the stronger this system: it gives a legal civil identity to everyone under its scope. 

The "first name + last name" couple is not, and never was, sufficient to identify someone without any ambiguity. Historians and anthropologists have often remarked that in small European villages, many individuals shared the same identity. In small setting where everyone was known to everyone, there was no "collective interest in the clear and unambiguous individuation of persons through their names" \cite{Pinacabral12}. In small villages nicknames (Big John), toponyms (John from the lake) and paraphrases (the son of Jake) could be much more efficient to distinguish someone from everyone else.

If this system worked for a long period of time, it was thanks to local agents of the state who could translate a local identity (Big John) into the civil identity needed by the state or the central authorities (John Martin) and reassure the state that  John Martin the conscript or John Martin the suspected tax evader was indeed Big John. With additional elements such as the precise date of birth, the place of birth, the names and profession of the parents... the first and last names could be used to identify someone in a much larger regional or national setting \cite{Noiriel01}. 

And today in our "global village" the first and last names are still the basis for worldwide identification. But without intimate knowledge or local agents in charge of the disambiguation, the collision of identities becomes problematic and more frequent. Every day in a random airport, someone sharing the identity of a known terrorist will be interrogated by customs agents or banned from flying. Someone will receive a parking ticket or a fine because she bears the same first and last name of someone else. Every second, bibliographic databases will try to differentiate John Lee the mathematician from John Lee the biologist in order to compute their scientific outputs \cite{Gomide17}. 

Yesterday's homonymy was the shared sign of belonging to the same locality. There may have been hundred of John Martins around 1700, but if they were not from the same place, they did not know they existed. Today's homonymy is shared between strangers in random places. In our interconnected societies, electronic social networks and  multiple registrations enable us to "meet" or to "bump into" people with the same names as ours, often in circumstances when we have to assert a right (to vote, to travel, to buy...) based on our civil identity. From the point of view of the individual, then, homonymy is a random annoyance, a discomfort or a personal catastrophe, depending on the circumstances.

But from the point of view of the manager of any large scale register, today's homonymy seems to be a very common nuisance, if we consider the great numbers of personal identifiers that are meant to distinguish individuals without ambiguity. Personal identification numbers such as the Social Security Number in the United States, or the "{\em numéro d'inscription au répertoire des personnes physiques}" (NIR) in France were created to resolve this particular problem \citeNP{MLLevy00}.

These numbers are not used daily by people who still prefer to be known by their names, and who do not gain anything by using a number instead. In the academic field, the "ORCID" promises to be "a persistent digital identifier that distinguishes you from every other researcher". It is meant to be used widely and the incentive is another promise : it "ensur[es] that your work is recognized".

But we do not know how frequent these identity collisions are. We do not know if, in a large scale society, many people have homonyms, or if only a small percentage does. 

This article\footnote{Additional material, including {\sf R} codes used for computations and to produce graphs, is available on a GitHub repository, {\sf https://github.com/freakonometrics/homonym}} brings forward an estimation of the proportion of homonyms in large scale groups based on the distribution of first names and last names in a subset of these groups. The estimation is based on the generalization of the "birthday paradox problem".

The main results is that, in societies such as France or the United States, identity collisions (based on first + last names) are frequent. The large majority of the population has at least one homonym. But in smaller settings, it is much less frequent : even if small groups of a few thousand people have at least one couple of homonyms, only a few individuals have an homonym.

%Consequences % in conclusion ??

\section{A Birthday Paradox Problem}

Consider a list of $k$ elements in $\mathcal{X}=\lbrace x_1,\cdots,x_k\rbrace$, and let us draw $n$ times, with replacement, so that $X_1,\cdots,X_n$ are i.id. random multinomial variables on $\mathcal{X}$ with probabilities $\boldsymbol{p}=(p_1,\cdots,p_k)$. In the birthday problem $\mathcal{X}$ are dates, $k=365$, and usually $\boldsymbol{p}$ is the uniform distribution on $\mathcal{X}$. One classical problem is to compute the probably to have (all) distinct values,
$$
\mathbb{P}[\forall j\neq i, X_i\neq X_j]
$$
or the proportion of observation with an {\em alter-ego}
$$
\frac{1}{n}\sum_{i=1}^n \boldsymbol{1}(\exists j\neq i,X_i=X_j).
$$
For the first problem, the birthday paradox is that when $n=25$ the first probability is close to $50\%$.

In this article, we will try to approximate the second one, interpreted as the proportion of people, within a group a size $n$, with an homonym. Let $Z_i$ denote the number of people (out of $n$) that share the same name with individual $i$,
$$
Z_i=\sum_{j\in\{1,\cdots,n\}\backslash\{i\}} \boldsymbol{1}(X_j \neq X_i)
$$
Hence, individual $i$ has an homonym if $Z_i\geq 1$. Thus, the proportion of people with an homonym is
$$
P_{n,k}=\frac{1}{n}\sum_{i=1}^n \boldsymbol{1}(Z_i\geq 0).
$$
If $\boldsymbol{p}$ is the uniform distribution on $\mathcal{X}$, then
$$
1-P_{n,k}=\binom{k}{n}\frac{1}{k^n}\sim \exp\left[
\binom{n+1}{2}\frac{1}{k}\right]\sim e^{-n^2/2k}
$$
see \citeNP{Chatterjeeetal04} for approximations and \citeNP{InoueAki08} and more recently \citeNP{CortinoBorja13} for surveys on computations of quantities related to the birthday problem.

\begin{figure}[!h]
    \centering
    \includegraphics[width=.24\textwidth]{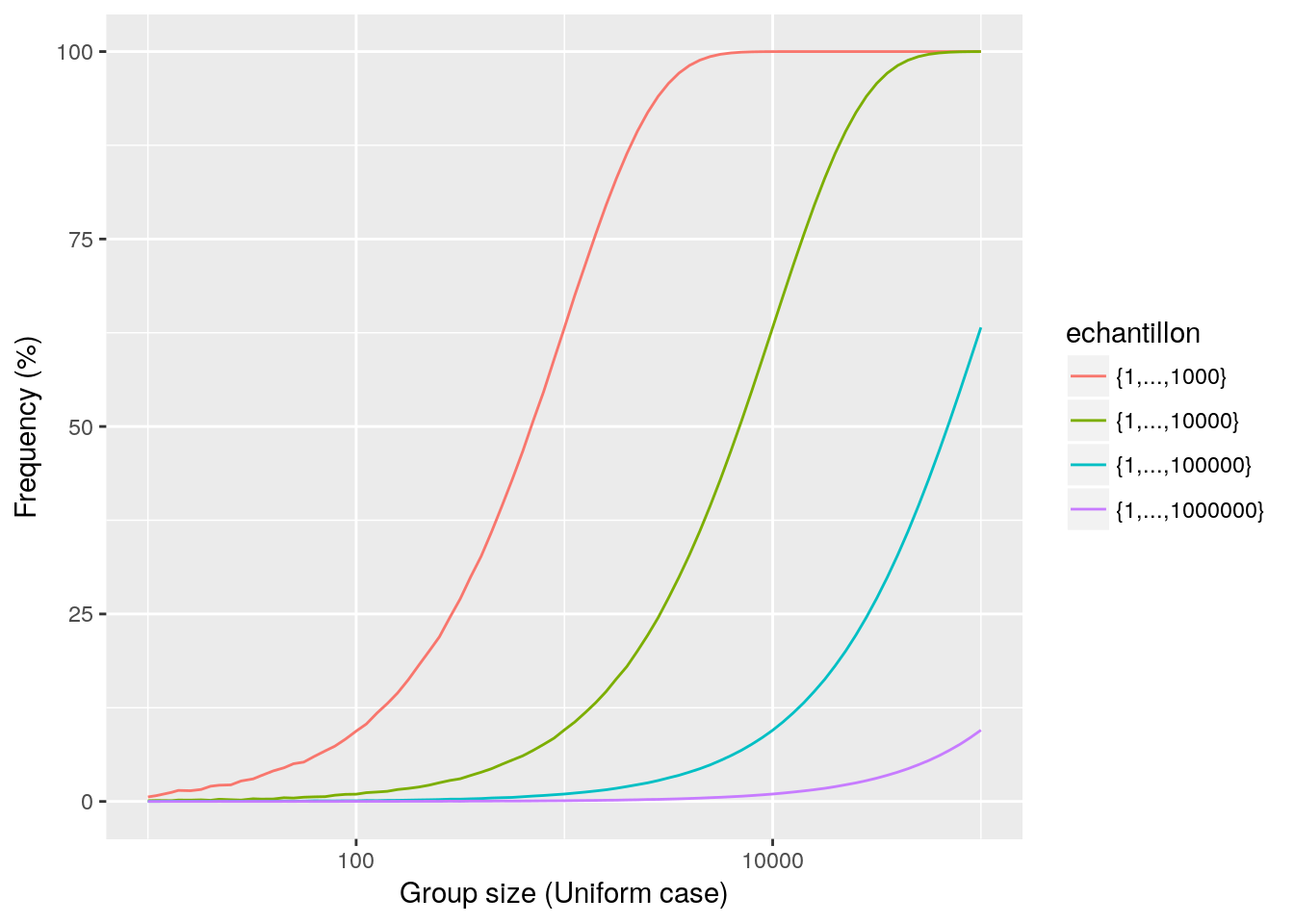}
    \includegraphics[width=.24\textwidth]{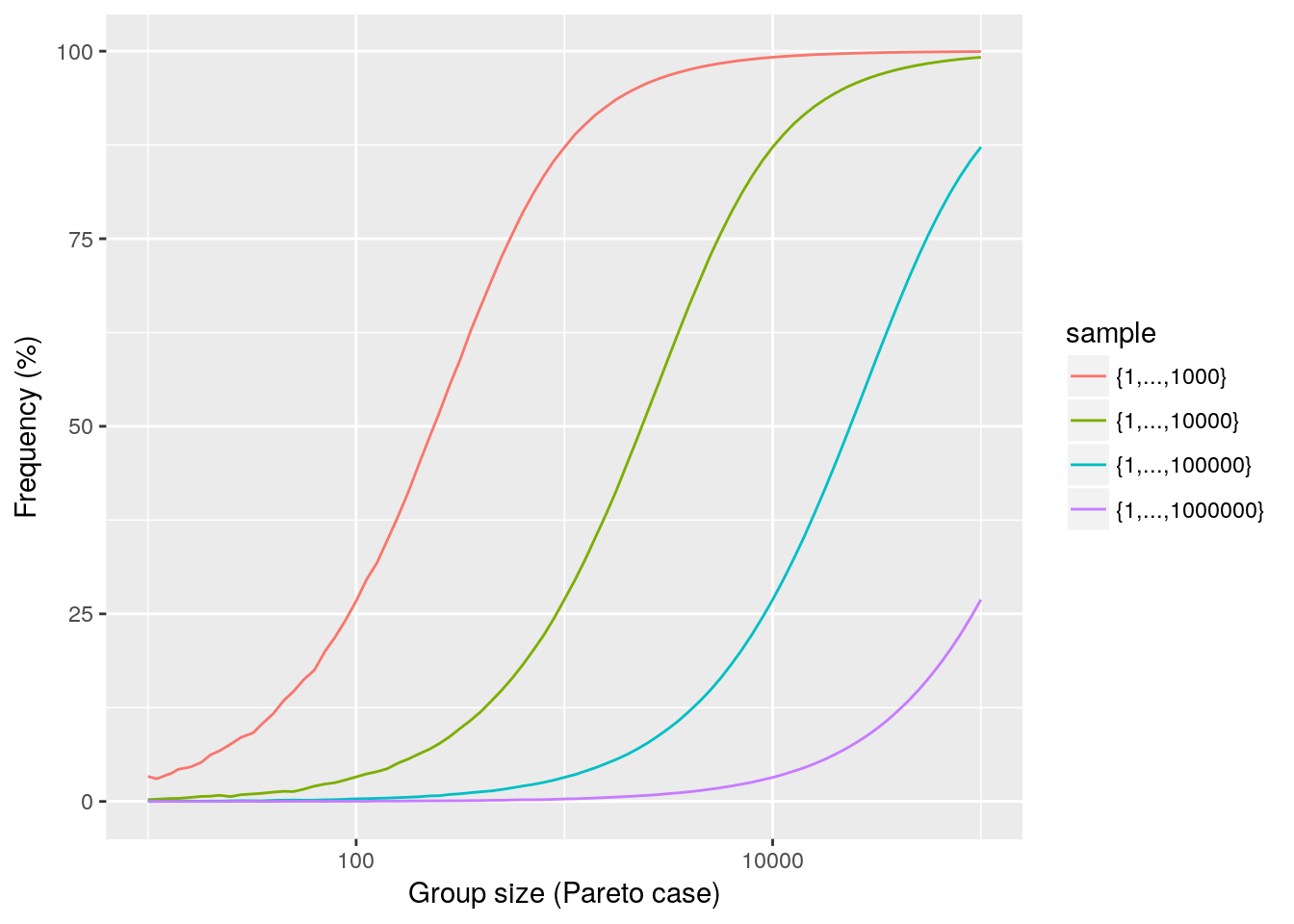}
    \includegraphics[width=.24\textwidth]{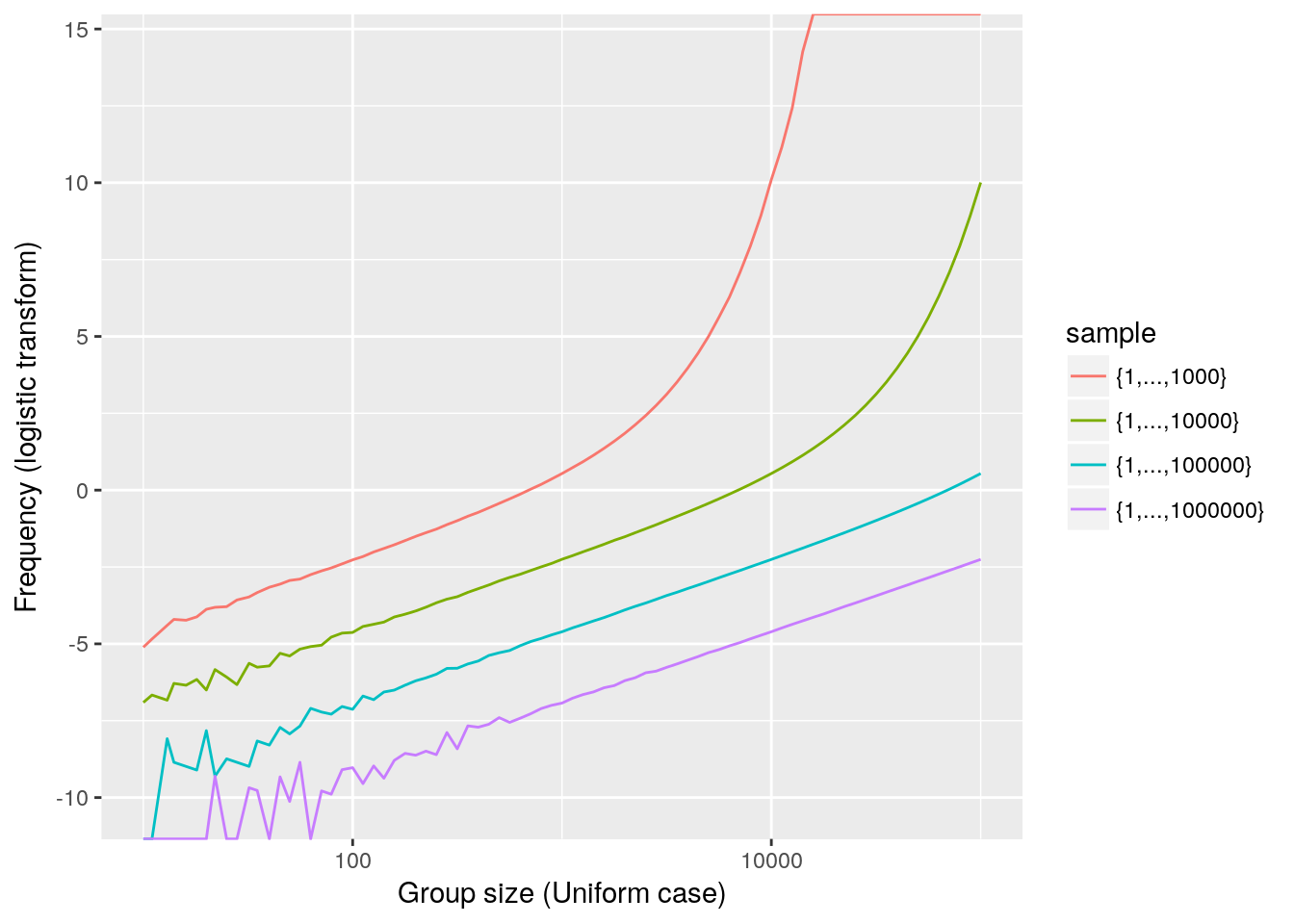}
    \includegraphics[width=.24\textwidth]{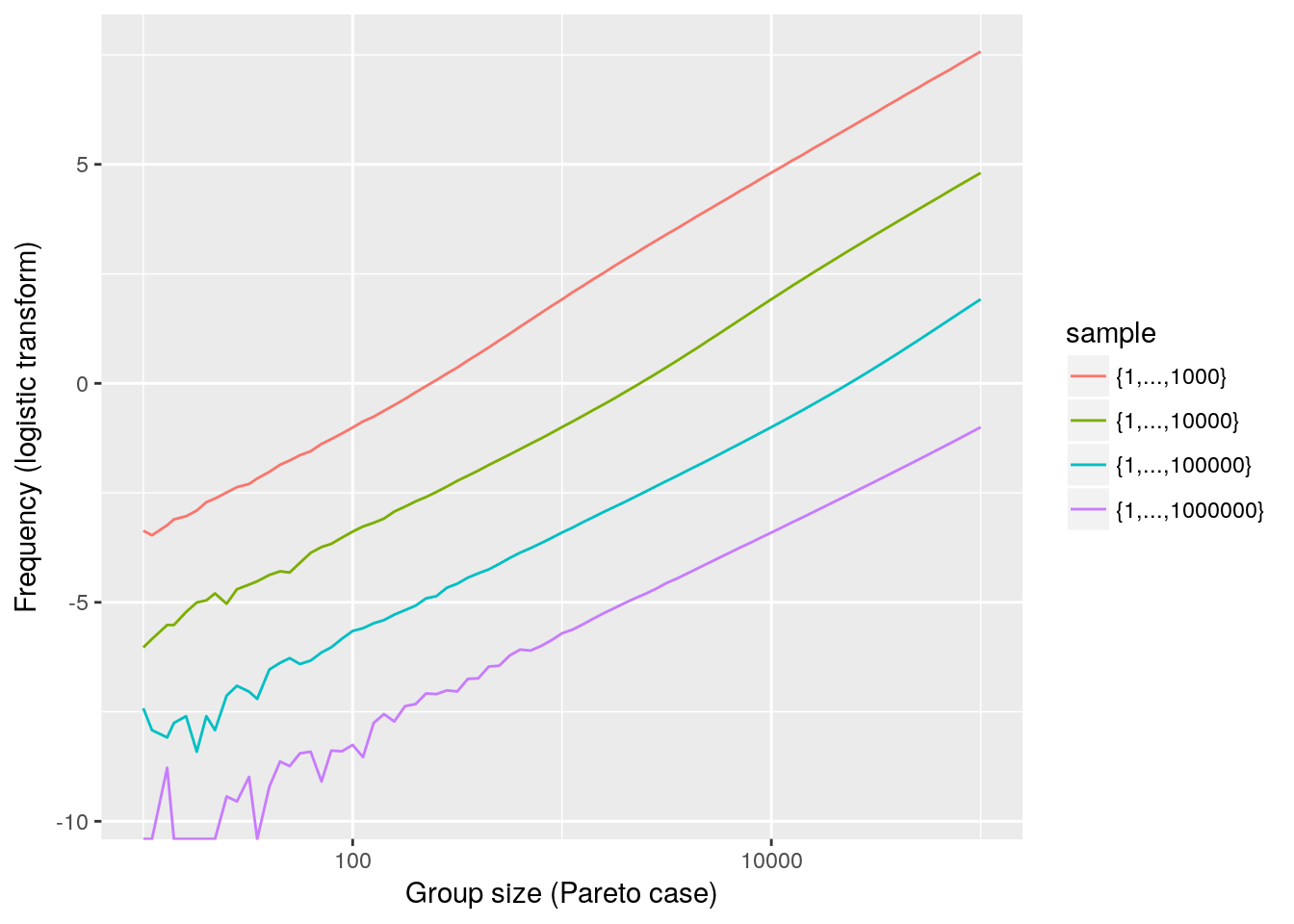}
    \includegraphics[width=.24\textwidth]{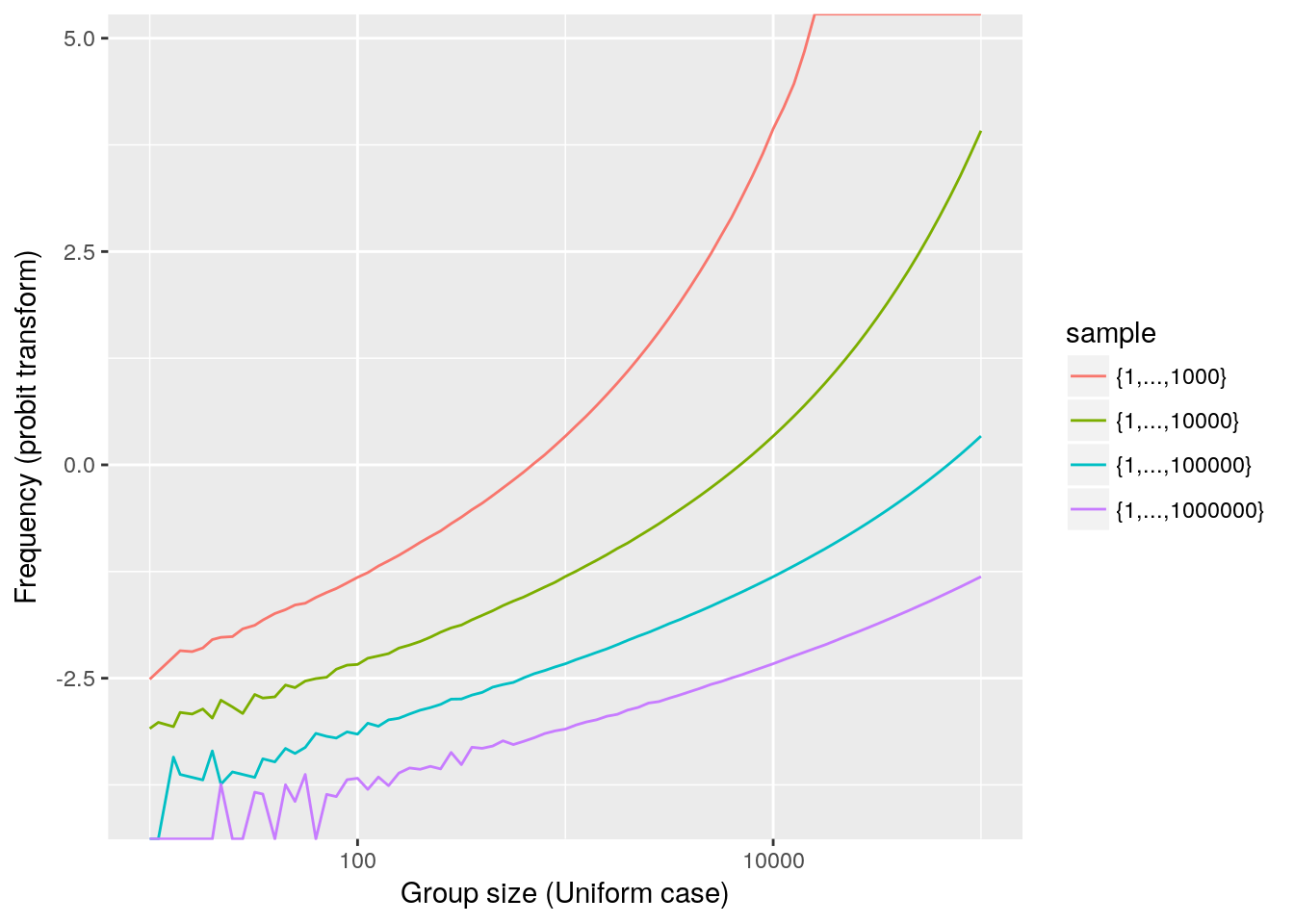}
    \includegraphics[width=.24\textwidth]{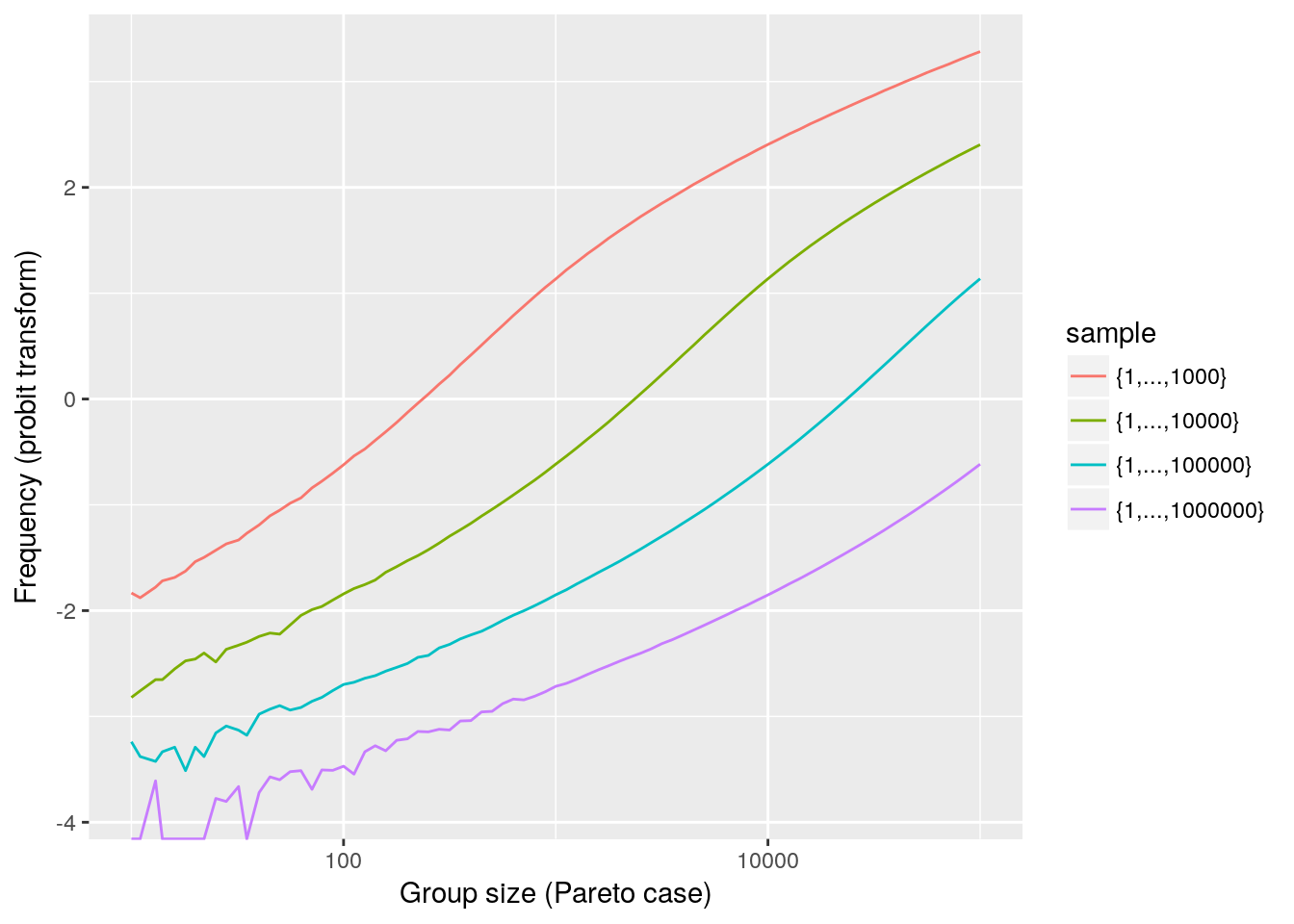}
    \caption{Evolution of $P_{n,k}$ as a function of the group size $n$ (on a log scale) as a function of $k$, for different distributions $\boldsymbol{p}$ (uniform on the left and Pareto on the right).}
    \label{fig:birthday0}
\end{figure}

Most properties derived analytically are based on the assumption that probabilities $\boldsymbol{p}$ are uniform. General properties are rather rare (see \citeNP{Munford77}, \citeNP{DasGupta}, \citeNP{InoueAki08}, or \citeNP{Nunnikhoven} for some attempts). From a numerical perspective, most quantities can be approximated using Monte Carlo simulations. Consider the case where of a set $\mathcal{X}$ (dates for the birthday problem, first or last names here) with size $k$, and consider a distribution $\boldsymbol{p}$ on $\mathcal{X}$. On Figure \ref{fig:birthday0} are computations of $P_{n,k}$ for various $k$ (the different lines on one graph), various $n$ (on the $x$-axis, with a log scale) and two specific distributions for $\boldsymbol{p}$: a uniform distribution on the left (as for the birthday problem) and a Pareto/Zipf law (closer to what can be observed on first names for instance, see \citeNP{Li} and the next section), where $p_i \propto i^{-\alpha}$.

\begin{figure}[!h]
    \centering
    \includegraphics[width=.24\textwidth]{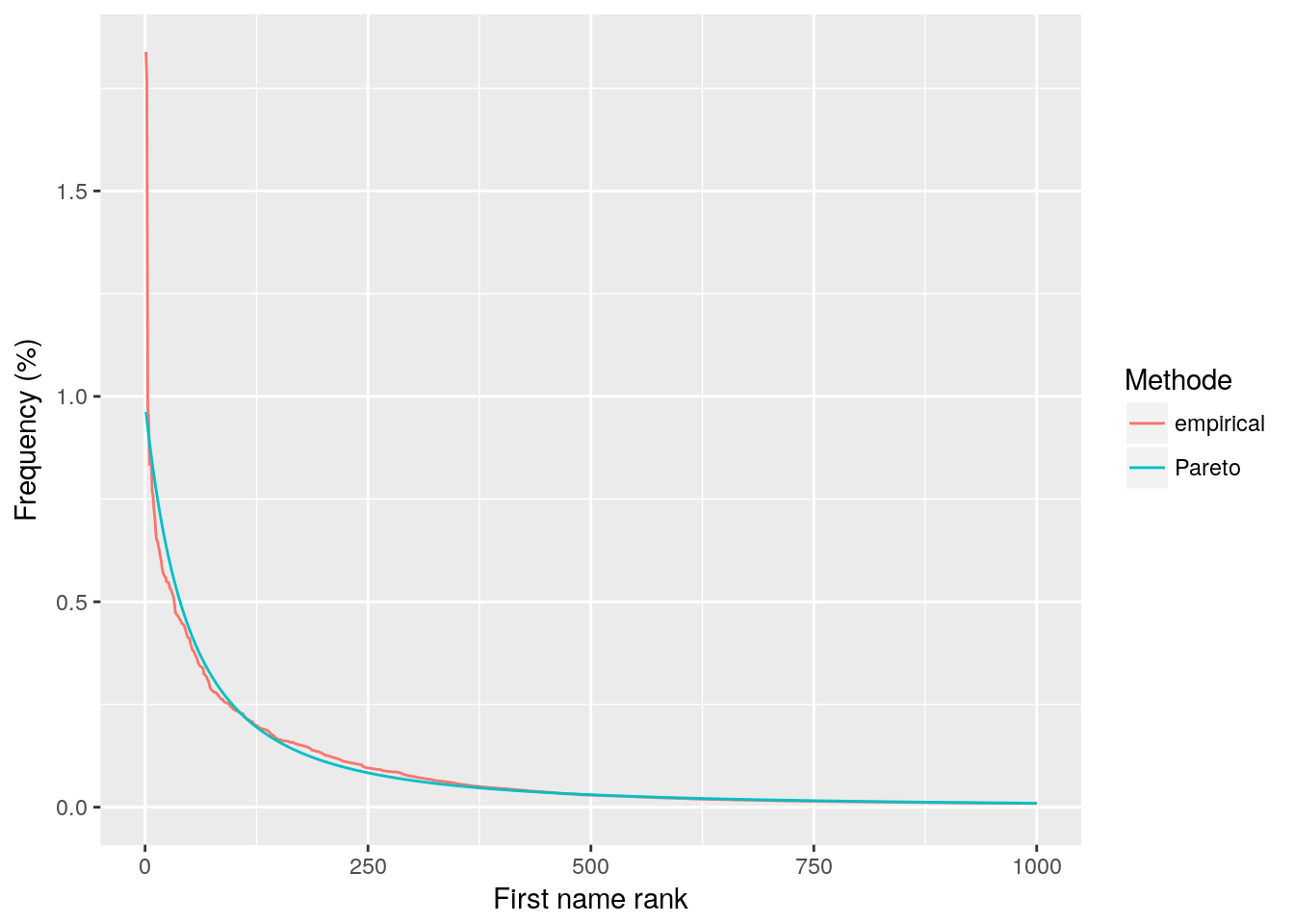}
    \includegraphics[width=.24\textwidth]{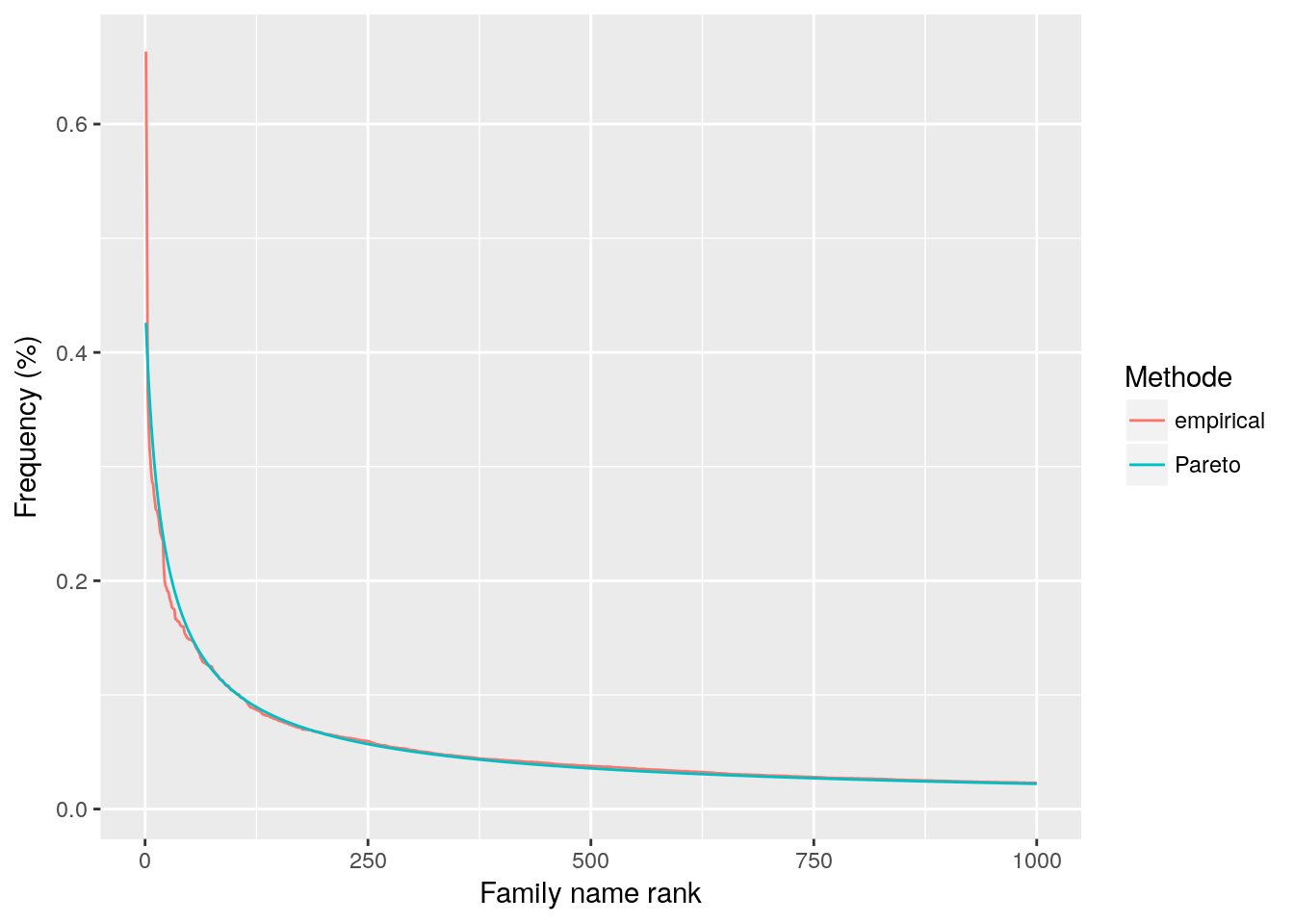}
    \caption{Empirical distribution of first (on the left) and last (on the right) names in France, with the estimate Pareto/Zipf fit.}
    \label{fig:birthday1}
\end{figure}

On Figure \ref{fig:birthday0}, $P_{n,k}$ is plotted on top and then two alternative graphs are presented: the evolution of $\text{logit}(P_{n,k})$ in the middle (where $\text{logit}(u)=\log[u/(1-u)]$) and $\Phi^{-1}(P_{n,k})$ below (which is the standard probit transform, where $\Phi$ denotes the cumulative distribution of the standard Gaussian centered distribution). Observe that with Pareto/Zipf distribution, a linear approximation can be considered,
$$
\Phi^{-1}(P_{n,k}) \sim a_{n,k}+b_{n,k} \log[n].
$$
And as show on Figure \ref{fig:birthday1}, on French first and last names, the Pareto/Zipf assumption is quite realistic. So linear approximations can be considered for transforms of $P_{n,k}$.

\section{First and Last Names}

In the case of homonyms, $\mathcal{X}$ is a set $\mathcal{X}_1\times\mathcal{X}_2$ since a person is characterized by a pair (first name, last name) with - potentially - $k_1$ first names and $k_2$ last names. With similar notations, let $P_{n,k_1,k_2}$ denote the proportion of homonyms, and let use Monte Carlo simulations to estimate that probability.

Let $\boldsymbol{p}=(p_{i_1,i_2})$ denote the empirical probability vector on $\mathcal{X}$. Note that on standard datasets, $\boldsymbol{p}$ contains a lot of zeros since many pairs have never been observed. For numerical simulation, let $\boldsymbol{p}^{\perp}$ denote the joint probability under the assumption that first and last names are independent,
$$
p^{\perp}_{i_1,i_2}=\left(\frac{1}{n}\sum_{j=1}^{k_2} p_{i_1,j}\right)\cdot\left(\frac{1}{n}\sum_{j=1}^{k_1} p_{j,i_2}\right)=
p_{i_1,\cdot}p_{\cdot,i_2}.
$$

\begin{figure}[!h]
    \centering
    \includegraphics[width=.4\textwidth]{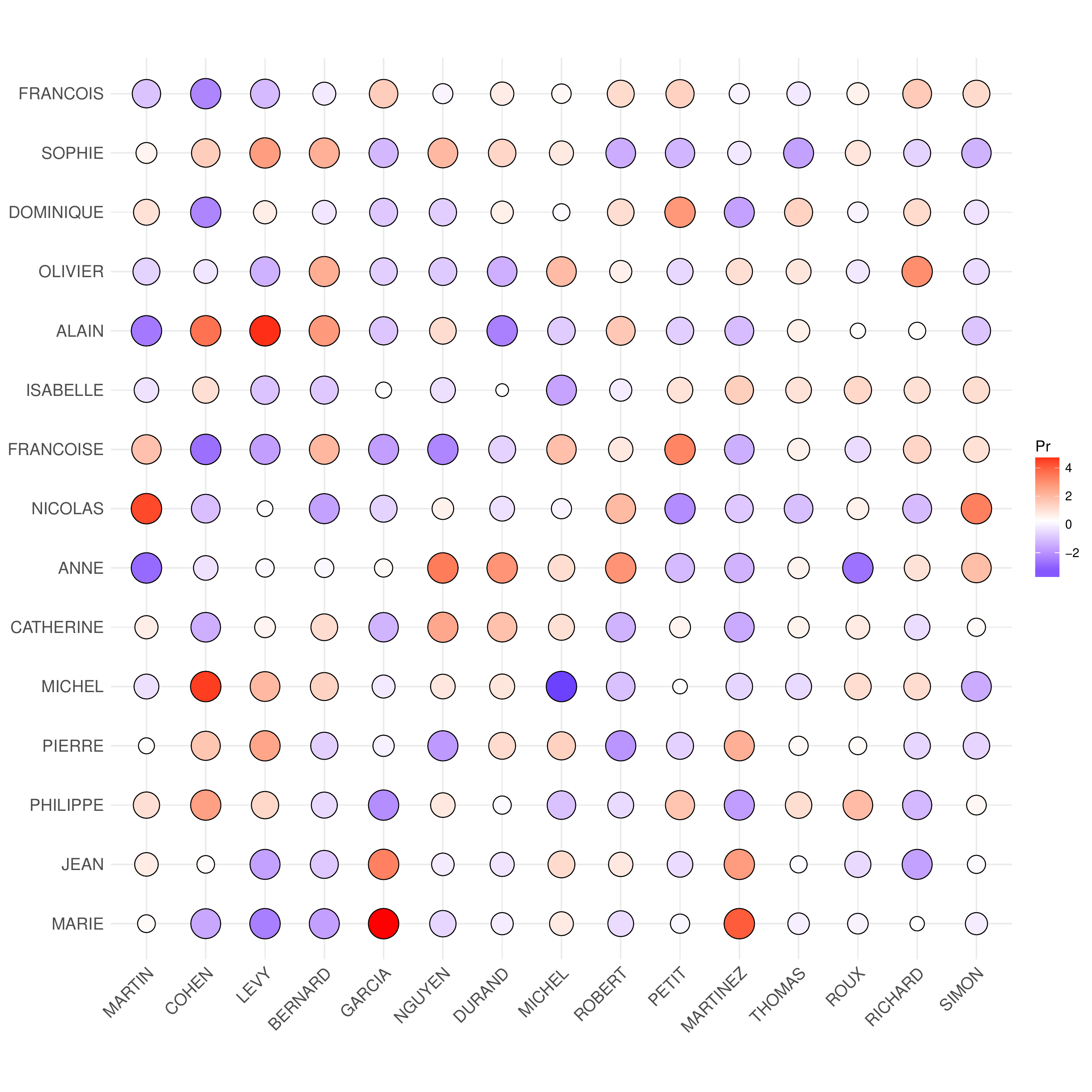}
    \caption{Pearson's residuals from a chi-square test of independence in the contingency table last vs. first names, in France..}
    \label{fig:birthday:joint}
\end{figure}

For Monte-Carlo simulation, we will draw $\boldsymbol{x}=(x_1,x_2)$'s according to either $\boldsymbol{p}$, or $\boldsymbol{p}^{\perp}$. As we can see on figure \ref{fig:birthday:joint}, drawing pairs according to $\boldsymbol{p}^{\perp}$ is not realistic, since first and last names are clearly not independent.

On the one hand, as we will see on two datasets, when drawing according to $\boldsymbol{p}^{\perp}$, the probit transformation of $P_{n,k_1,k_2}$ is linear in $n$. On the other hand, when drawing from $\boldsymbol{p}$, we underestimate the true probability when $n$ is too large, if we use (for $\boldsymbol{p}$) empirical frequencies on a too small dataset. Nevertheless, it is possible to fit a linear model when $n$ is not too small, in the later case, and then extrapolate it.

%\citeNP{Li} 

\section{Application on French Data}

\begin{figure}[!h]
    \centering
    \includegraphics[width=.24\textwidth]{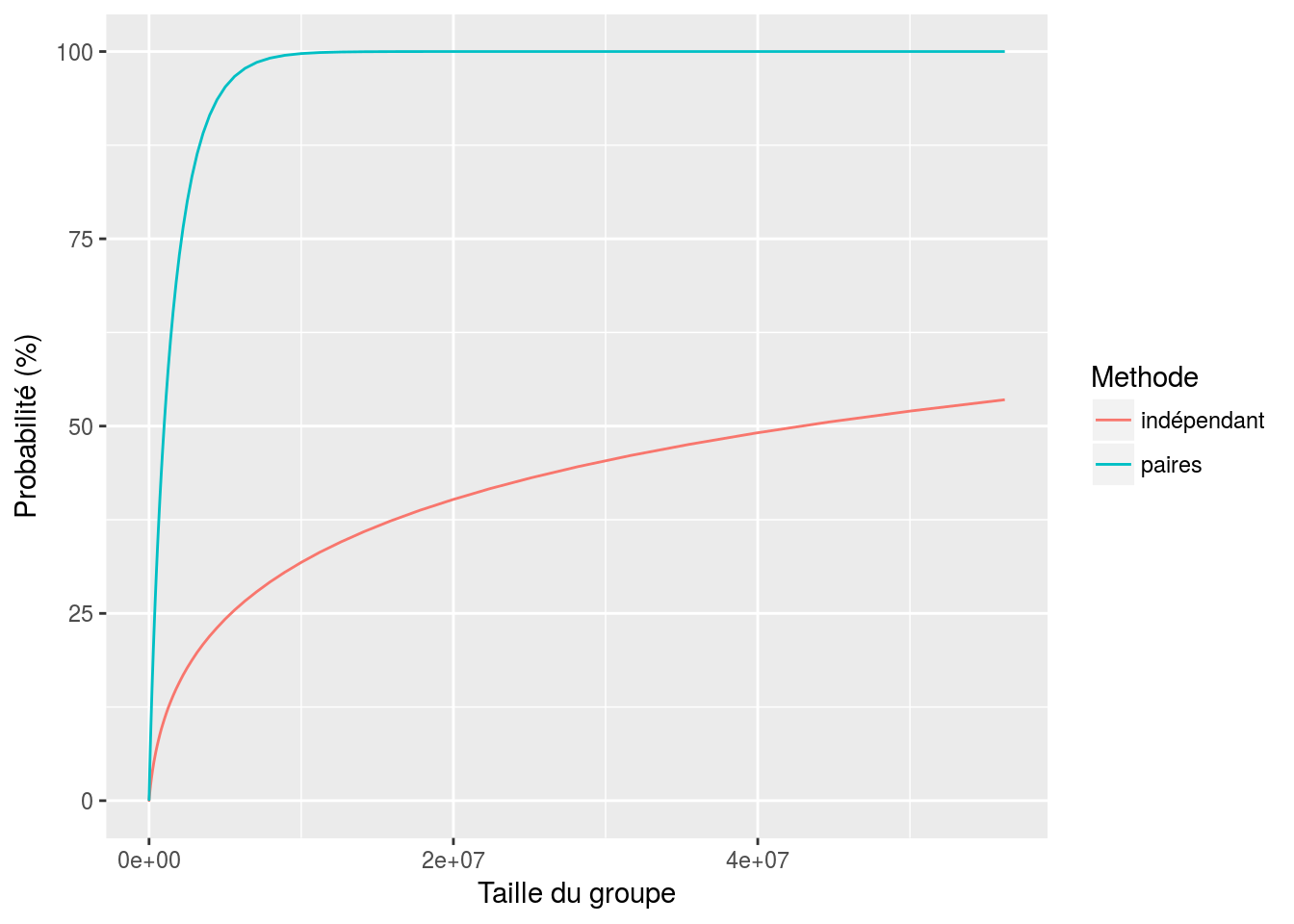}
    \includegraphics[width=.24\textwidth]{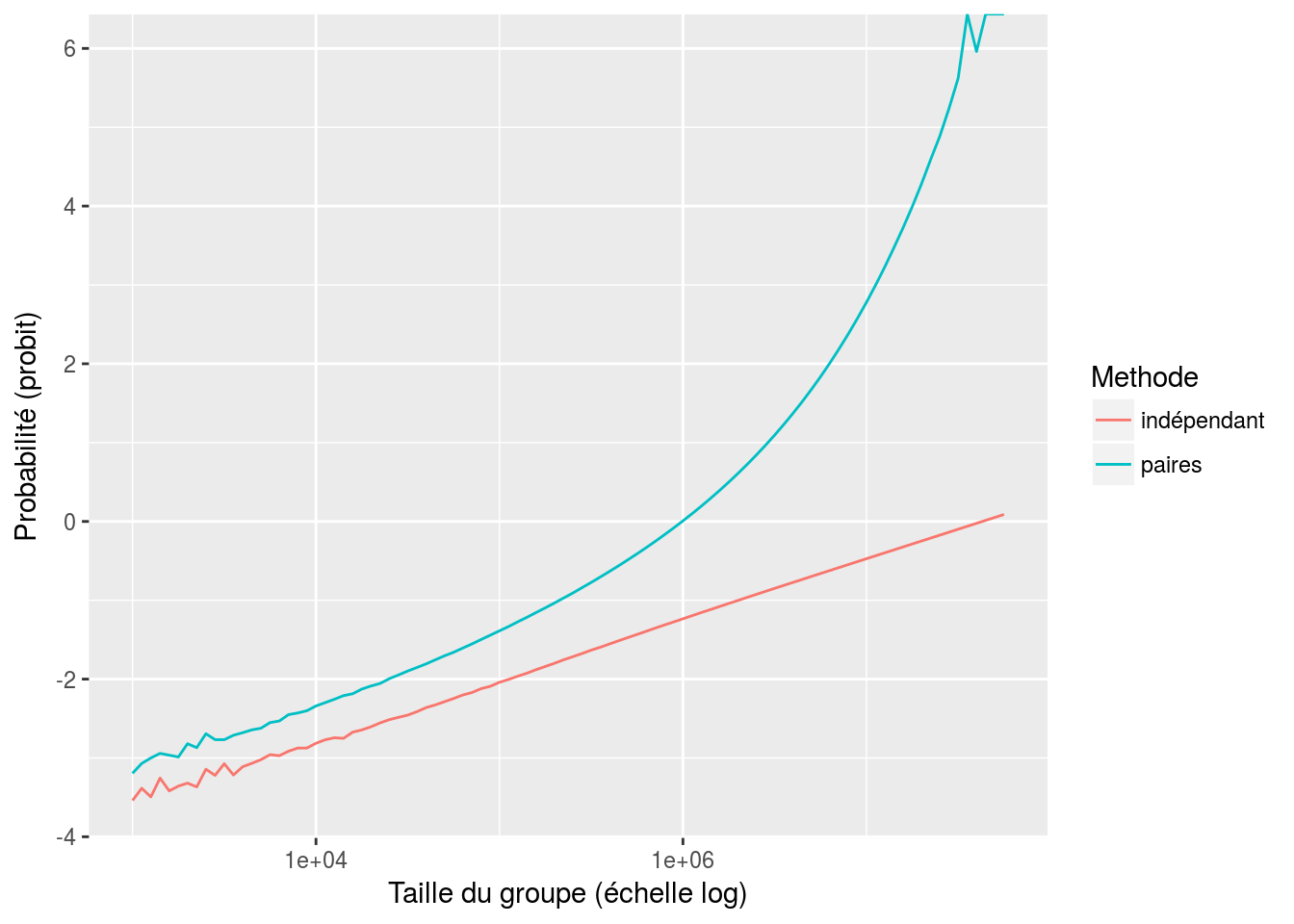}
    \caption{Proportion of homonyms when drawing from the French population in Paris, either drawing pairs (first and last names) in blue, or drawing first and last names independently, in red. Empirical probabilities $P_n$ are on the left, and the probit transform $\Phi^{-1}(P_n)$ is on the right.}
    \label{fig:birthday2}
\end{figure}

In order to compute the probability $P_n$ in the context of French names, the electoral roll of Paris and Marseille (for the year 2015) has been used. In this dataset, we have the first name, last name and date of birth of registered electors in Paris (1,757,895 observations)
Overall, we kept 1,542,528 observations, because of some typo in the original dataset. There where $k_1=74,085$ first names in that dataset, and $k_2=309,907$ last names (actually almost half of those appeared only once). Because of the variety of the first and last names, our sample size ($n=1.5$ million) was too small to estimate the proportion of people with an homonym in the entire French population ($65$ million). Resampling from pairs (first and last names together) will over-estimate the proportion of homonyms on a very large group. Nevertheless, as mentioned in the previous section, drawing independently first and last names is not realistic, since both are clearly correlated. On Figure \ref{fig:birthday2} we can visualize the proportion of homonyms when drawing from the French population in Paris, either drawing pairs or drawing first and last names independently.

\begin{figure}[!h]
    \centering
    \includegraphics[width=.24\textwidth]{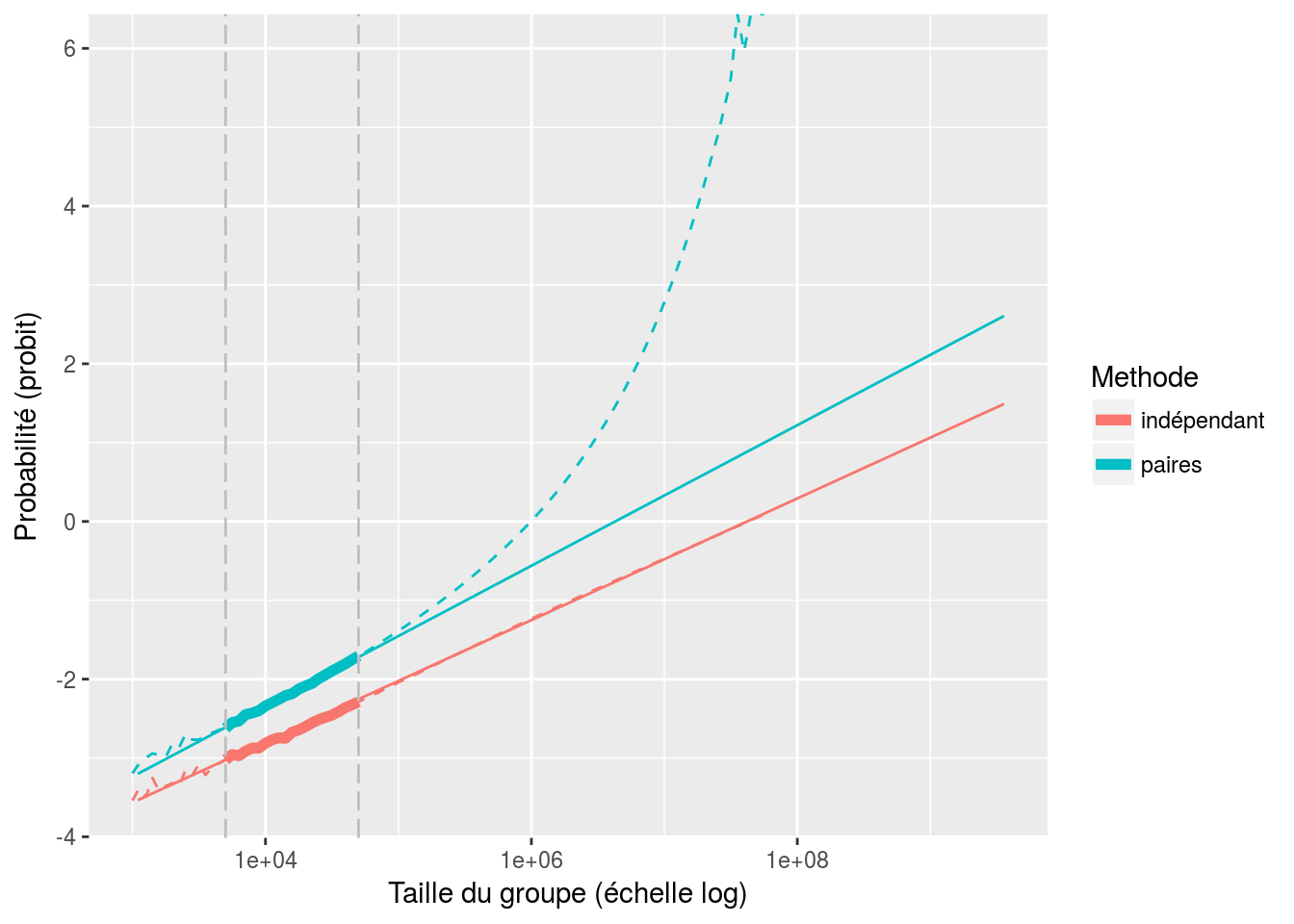}
    \includegraphics[width=.24\textwidth]{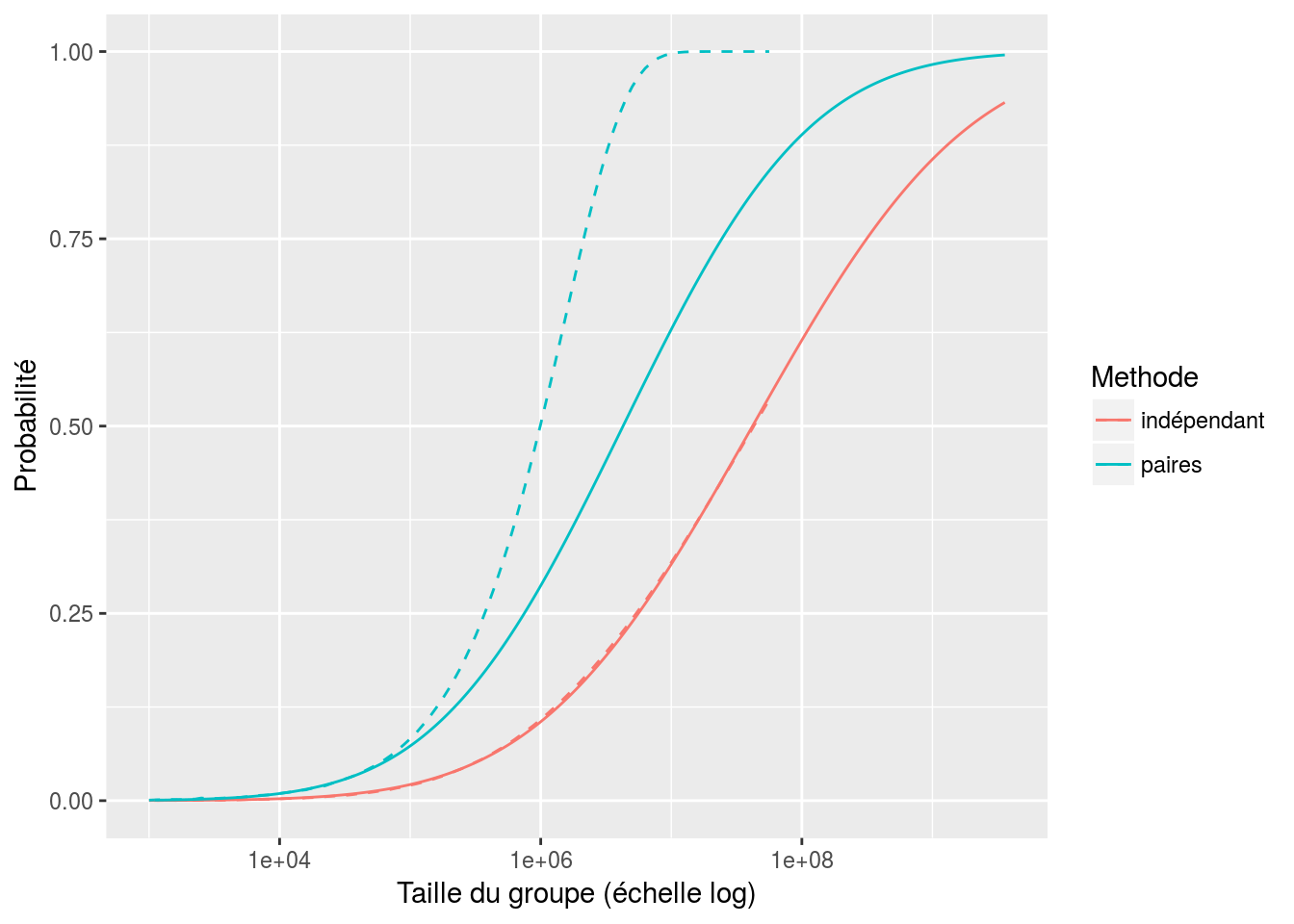}
    \caption{Proportion of homonyms with a linear extrapolation when pairs were drawn (linear on the {\em probit} transform as a function of $\log n$).}
    \label{fig:birthday3}
\end{figure}

When $n$ is not two large, drawing pairs should yield a good approximation. On Figure \ref{fig:birthday3} we use a linear approximation when $n$ is between 5,000 and 50,000. Then we extrapolate that linear approximation for large $n$'s. Hence, 

\section{Temporal evolution of that Proportion}

On two larger datasets\footnote{The first one is the {\em fichier des prénoms, 2016 edition} available from {\sffamily https://www.data.gouv.fr/fr/datasets/fichier-des-prenoms-edition-2016/} produced by the National Institute of Statistics (INSEE) and the second one is the {\em fichiers des noms de famille - 1891-1990 - 1999 edition}, produced by INSEE, available from ADISP-CMH.}, we can observe the evolution of first and last names in France, see Tables \ref{tab:birthday1} and \ref{tab:birthday2} (those datasets contained statistics about first and last names, respectively, but not paired).

\begin{table}[!h]
    \centering
\begin{tabular}{|lrrr|}\hline
time period &size & top 10 & top 100 \\ \hline
1916-1940& 95,000 & $25.17\%$ & $79.05\%$ \\
1941-1965& 105,000 & $20.50\%$ &$72.61\%$\\
1966-1990& 245,000 & $12.59\%$ & $56.98\%$ \\ \hline
\end{tabular}
\caption{First names in France.}
    \label{tab:birthday1}
\end{table}

\begin{table}[!h]
    \centering
\begin{tabular}{|lrrr|}\hline
time period &size & top 10 & top 100 \\ \hline
1916-1940 & 638,000&$1.83\%$&$8.66\%$ \\
1941-1965& 669,000&$1.76\%$&$8.41\%$\\
1966-1990& 814,000&$1.57\%$&$7.83\%$\\ \hline
\end{tabular}
\caption{Last names in France.}
    \label{tab:birthday2}
\end{table}

\begin{figure}[!h]
    \centering
    \includegraphics[width=.4\textwidth]{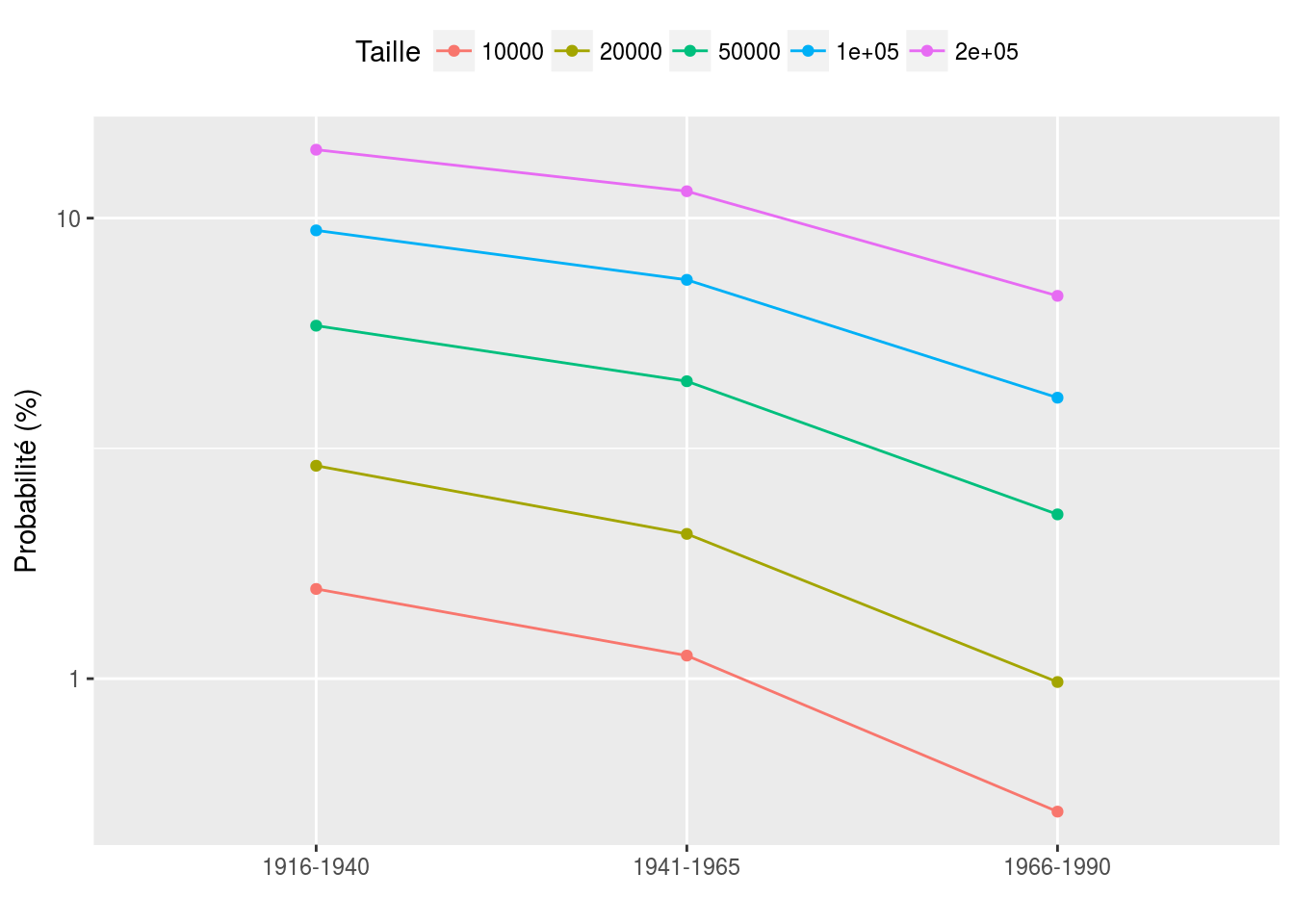}
    \caption{Evolution of the proportion of homonyms, $P_n^\perp$, assuming independence between first and last names, .}
    \label{fig:birthday4}
\end{figure}

It is then possible, assuming independence between first and last names, to visualize the evolution of the proportion of homonyms, approximated using Monte Carlo simulations, on Figure \ref{fig:birthday4}, for goupes of size 10,000 up to 200,0000 people (from bottom to top).  

\section{Application to Ohio Data}

\begin{figure}[!h]
    \centering
    \includegraphics[width=.24\textwidth]{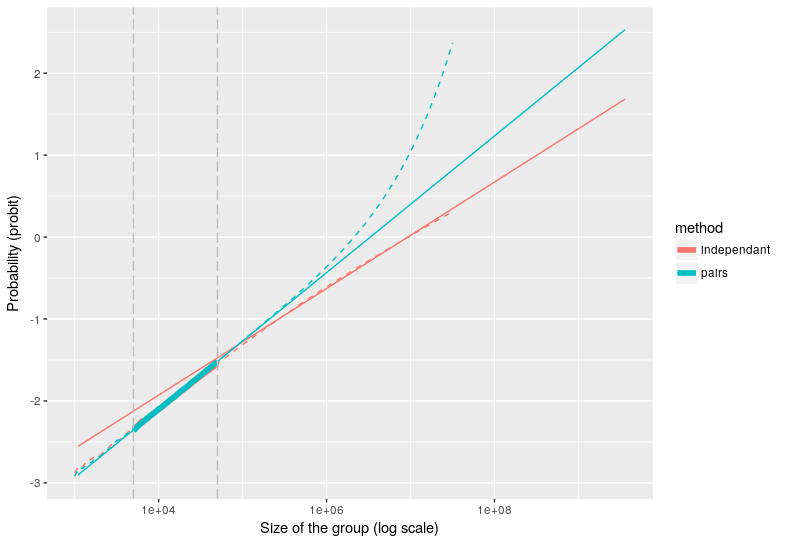}
    \includegraphics[width=.24\textwidth]{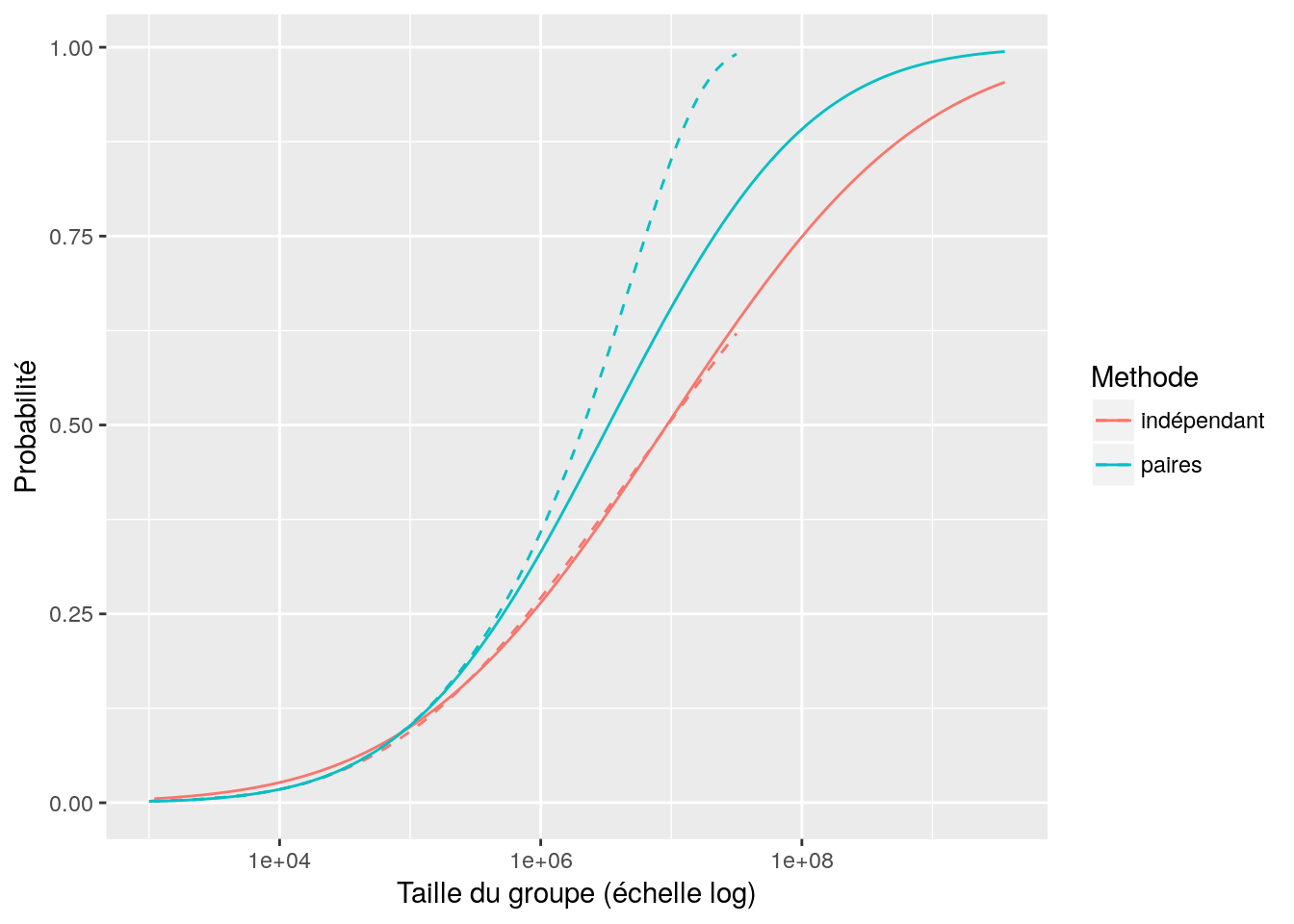}
    \caption{Proportion of people having an homonym is a group of size $n$, in Ohio.}
    \label{fig:birthday7}
\end{figure}

In order to see how general our technique is, consider the dataset of voters in Ohio, in the United-States. It is a dataset with 7.8 million individuals. Observe that in that group, 50\% people have an homonym in that specific state. It might be interesting to extrapolate to a much higher $n$. As described in Figure \ref{fig:birthday7}, in a population of $n=320$ million people, we can estimate that $95.1\%$ have an homonym.

\section{Conclusion}

As the interconnexion of our world increases and as the realm of interactions widens , we encounter a increasing number of homonyms. These collisions are annoying. But we continue to value the use of a basic identification system. Some contemporary changes reduce the chance of collisions: we increasingly choose rare names for our children, and, at least in Europe, the transmission of the father's last name is slowly replaced by the possibility to choose to transmit the mother name or to create a combination of both parents' names. %see for example http://europa.eu/rapid/press-release_IP-12-644_fr.htm?locale=FR or the French law of 2003 

\section{Datasets}

\begin{itemize}

\item Ohio Voter Files available at https://www6.sos.state.oh.us/ords/f?p=111:1

\item Paris and Marseille Voter Files 

\item Fichier des prénoms, édition 2016, INSEE, available at https://www.data.gouv.fr/fr/datasets/fichier-des-prenoms-edition-2016/

\item Fichiers des noms de famille - 1891-1990 - Édition 1999, INSEE [producteur], ADISP-CMH [diffuseur] 
\end{itemize}

\end{document}